\documentclass[sigconf]{acmart}
\usepackage{enumitem}
\usepackage{algorithm}
\usepackage{algorithmic}
\usepackage{multirow}
\AtBeginDocument{%
  }

\copyrightyear{2026}
\acmYear{2026}
\setcopyright{cc}
\setcctype{by-nc-nd}
\acmConference[KDD '26]{Proceedings of the 32nd ACM SIGKDD Conference on Knowledge Discovery and Data Mining V.1}{August 09--13, 2026}{Jeju Island, Republic of Korea}
\acmBooktitle{Proceedings of the 32nd ACM SIGKDD Conference on Knowledge Discovery and Data Mining V.1 (KDD '26), August 09--13, 2026, Jeju Island, Republic of Korea}
\acmPrice{}
\acmDOI{10.1145/3770854.3780278}
\acmISBN{979-8-4007-2258-5/2026/08}

\begin{document}

\title{KFTD: Koopman-Fourier Time-Differentiable Network for Continuous Ocean Spatiotemporal Forecasting}

\author{Qinghui Chen}
\orcid{0000-0001-5504-2951}
\affiliation{%
  \institution{Shandong University}
  \department{School of Control Science and Engineering}
  \city{Jinan}
  \country{China}
}
\affiliation{%
  \institution{Laoshan Laboratory}
  \city{Qingdao}
  \country{China}
}\email{202420785@mail.sdu.edu.cn}

\author{Zekai Zhang}
\orcid{0000-0002-1312-5828}
\affiliation{%
\institution{Shandong University}
\department{School of Control Science and Engineering}
  \city{Jinan}
  \country{China}}
\email{202420810@mail.sdu.edu.cn}

\author{Hailong Liu}
\orcid{0000-0002-8780-0398}
\affiliation{%
 \institution{Chinese Academy of sciences}
 \department{Institute of Atmospheric Physics}
 \city{Beijing}
 \country{China}}
 \email{lhl@lasg.iap.ac.cn}

\author{Jinglin Zhang}
\authornote{Corresponding author}
\orcid{0000-0003-1618-8493}
\affiliation{%
  \institution{Shandong University}
  \department{School of Control Science and Engineering}
  \city{Jinan}
  \country{China}}
  \email{jinglin.zhang@sdu.edu.cn}

\author{Cong Bai}
\orcid{0000-0002-6177-3862}
\affiliation{%
  \institution{Zhejiang University of Technology}
  \department{College of Computer Science}
  \city{Hangzhou}
  \country{China}}
\email{congbai@zjut.edu.cn}

\renewcommand{\shortauthors}{Qinghui Chen et al.}

\begin{abstract}
Accurate oceanic forecasting is critical for climate monitoring and disaster early-warning. However, ocean spatiotemporal forecasting encounters the double challenges of modeling complex dynamical systems and ensuring computational efficiency. We present Koopman–Fourier Time-Differentiable (KFTD) Network, a time-continuous two-stage paradigm that decouples interpolation from prediction to achieve efficient and scalable spatiotemporal modeling. We map complex nonlinear dynamics into the Koopman linear space and exploit Fourier analysis to enable continuous-time interpolation at arbitrary sub-steps.  A lightweight residual network consumes the high-fidelity intermediate states to yield the final forecast.  Unlike diffusion models, KFTD eliminates multi-step noise sampling and directly evolves the system in continuous time, yielding a 4× computational speed-up.  We further introduce a D-PP Loss that supports arbitrary PDE constraints in an end-to-end manner, breaking the physical-consistency bottleneck of pure data-driven approaches. Empirical results on four ocean datasets confirm that our continuous-time framework reduces MSE by an average of 5.6\% (up to 12.7\% for SST) and improves efficiency over MCVD by 76.25\%.
\end{abstract}

\ccsdesc[500]{Computing methodologies~Temporal reasoning}
\ccsdesc[300]{Computing methodologies~Spatial and physical reasoning}
\ccsdesc[300]{Applied computing~Environmental sciences}

\keywords{Spatiotemporal Forecasting, Ocean Dynamics, Operational Ocean Forecasting, Physics-Informed Learning, Koopman Neural Operator}


\maketitle

\newcommand\kddavailabilityurl{https://doi.org/10.5281/zenodo.18054522}
\ifdefempty{\kddavailabilityurl}{}{
\begingroup\small\noindent\raggedright\textbf{Resource Availability:}\\

The source code of this paper has been made publicly available at \url{\kddavailabilityurl}.
\endgroup
}

\section{Introduction}

\begin{figure}
\centering
    
\includegraphics[width=1\linewidth]{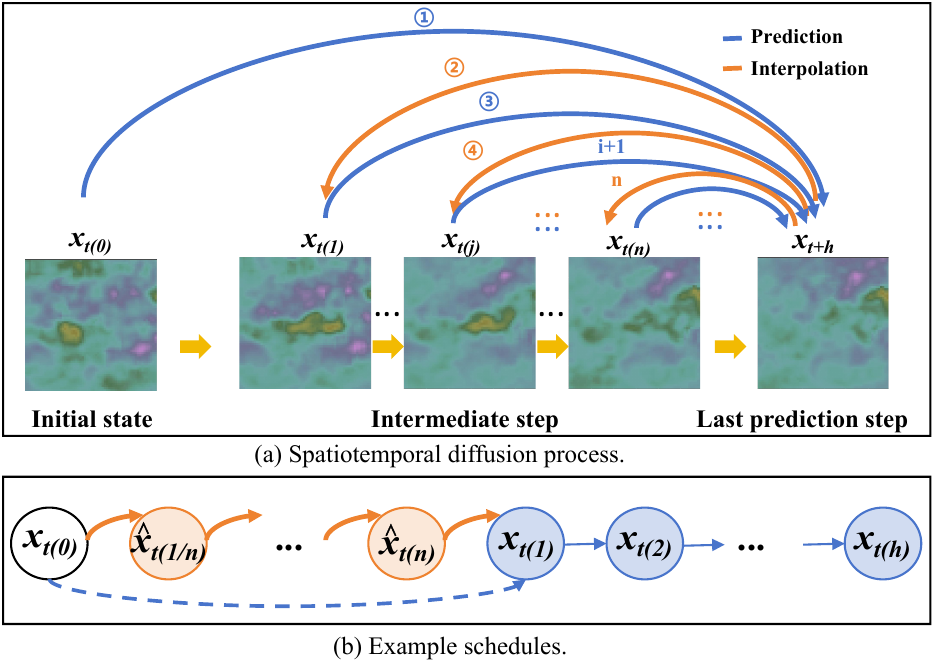}

\caption{Time-differentiable two-stage framework. (a) Initiate the forecast with initial conditions; predict $x_{t+h}$ using the prediction network; interpolate intermediate steps with the interpolation network. Repeat to cover the full forecast horizon. (b) Uniformly map $n$ time-differentiable steps between input $x_0$ and output $x_1$.}
\label{fig1}
\end{figure}
Oceans control Earth’s energy, water, and biogeochemical cycles, modulating weather patterns from tropical cyclones to multi-decadal climate oscillations. The ability to forecast oceanic state variables — spanning temperature, salinity, velocity, dissolved oxygen, carbon content, and biogeochemical tracers—at lead times from hours to seasons underpins early-warning systems for marine heatwaves, harmful algal blooms, storm surges, and global climate extremes.

This forecasting task is uniquely demanding. The governing fluid-dynamics equations are nonlinear, multi-scale (from millimetre turbulence to planetary waves), and computationally intractable at operational resolutions; observational data remain sparse and intermittently noisy; and stringent latency requirements preclude the brute-force numerical integration traditionally used in weather and climate centres ~\cite{31liu2024spatiotemporal,41verezemskaya2021assessing,42pontoh2024deep,43liu2024predictability,huang2025benchmark}. The data-driven alternatives leverage deep neural networks to learn spatiotemporal correlations directly from observations~\cite{6stofa2020deep,7feng2021study,8zhang2022oceanic}.  Once trained, such models yield rapid inference~\cite{34li2023astmen}, but they generalise poorly under sparse or noisy observations.

Diffusion forecasters discretise prediction into many denoising steps, ~\cite{15xing2024survey,16yang2023diffusion,18ruhling2024dyffusion,35xu2021spatiotemporal}. But forcing either excessive computational overhead for fine resolution or aliasing of high-frequency transients~\cite{17feng2023diffpose,33gou2020deepocean}. The need for tens to hundreds of neural function evaluations per forecast horizon that is unacceptable for operational deployment.

Achieving reliable ocean forecasts demands simultaneous fidelity to the underlying geophysical laws expressed by partial differential equations (PDEs) and computational efficiency sufficient to deliver timely guidance at high spatiotemporal resolution. Classical numerical ocean models solve discretised PDEs with finite-difference or finite-volume schemes~\cite{32o2022spatio,44raissi2019physics,45codiga2011unified}.  While physically consistent, these solvers incur prohibitive cost when resolving sub-mesoscale processes and are sensitive to the availability and quality of initial and boundary conditions~\cite{3li2010application,4wu2016improving,5li2019simple}.

In summary, current spatiotemporal ocean forecasting is hindered by three intertwined challenges: \textbf{(i)} the computational and representational limitations of discrete diffusion models, whose fixed temporal grids and multi-step noise sampling introduce aliasing of high-frequency transients and latency incompatible with operational deadlines; \textbf{(ii)} the modeling of strongly nonlinear, multi-scale dynamics spanning from millimetre turbulence to planetary waves, where conventional approaches struggle to reconcile global linearization with localized nonlinear perturbations; and \textbf{(iii)} the physical-inconsistency bottleneck of purely data-driven schemes that, under sparse or noisy observations, frequently violate fundamental conservation laws—mass, momentum, and energy—yielding geophysically implausible predictions.

We present KFTD (Koopman–Fourier Time- Differentiable Network), a time-continuous two-stage paradigm. Resolves these limitations explicitly \emph{decoupling temporal interpolation from spatiotemporal prediction}.  
As illustrated in Figure~\ref{fig1}(a), a lightweight prediction network advances the system state over a coarse horizon \(t \to t+h\).  A second, fully differentiable interpolation network then analytically generates any desired sub-step within \([t,t+h]\) by evolving the latent trajectory in the linear Koopman space via Fourier operators.  The entire pipeline is trained end-to-end without Monte-Carlo sampling, yielding a single forward pass for arbitrarily dense forecasts.  Consequently, KFTD eliminates the multi-step noise sampling endemic to diffusion models. To enforce physical consistency, we introduce a modular Data-Physics Prior (D-PP) that imposes arbitrary PDE residuals in an end-to-end manner.  Unlike penalty-based approaches that require hand-tuned weighting, our loss is additive and modular: any additional conservation law can be incorporated by simply appending its residual term without re-engineering the architecture.

Our contributions can be summarized as follows:

\textbf{(i) Framework:} We introduce KFTD, a time-differentiable two-stage framework that decouples continuous-time interpolation from prediction, superseding discrete diffusion paradigms.

\textbf{(ii) Mechanism:} We introduce depth-adaptive Fourier coefficient learning within a multiscale Koopman decomposition, surpassing fixed-frequency limits to extract multiscale periodic features adaptively.

\textbf{(iii) Constraint:} A Modular Physics-Consistent D-PP Loss is provided for plug-and-play enforcement of any PDE, ensuring end-to-end accuracy and physical consistency.

Through empirical testing and validation of historical data, we determined that the proposed model outperforms known prediction methods in accuracy. The Code will be made public upon publication.

\section{Related work}

\begin{figure*}
    \centering
    \includegraphics[width=1\linewidth]{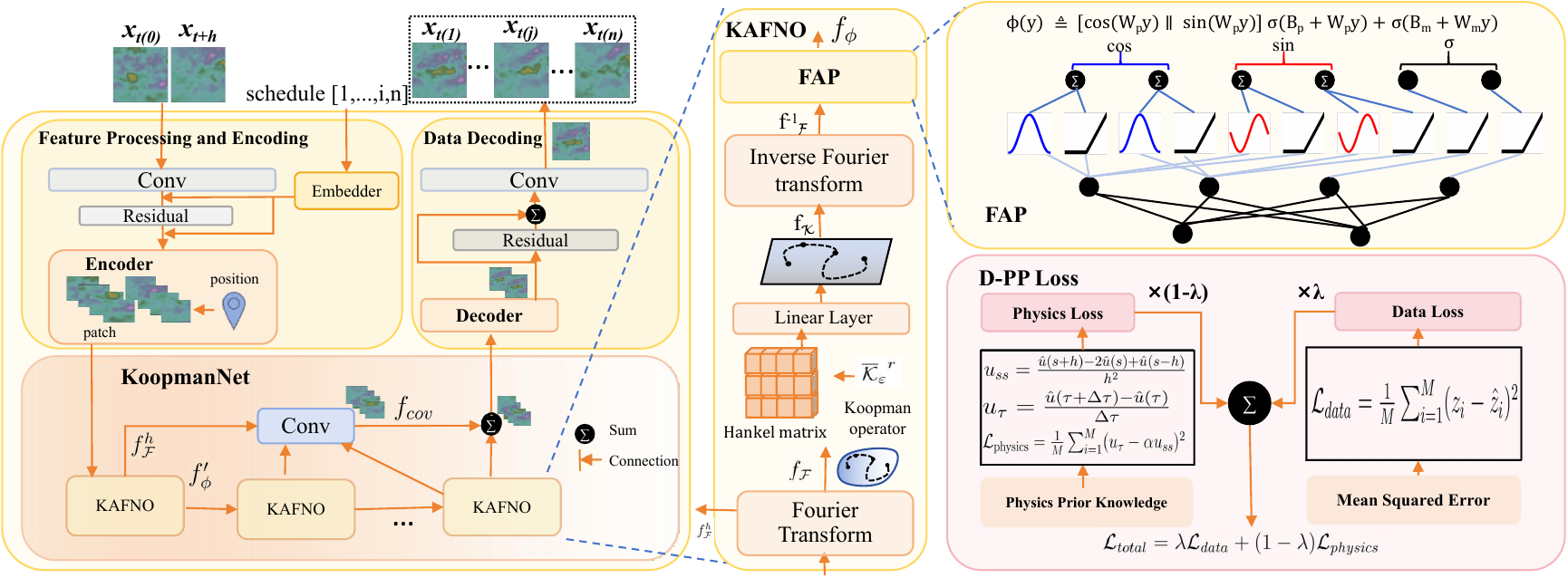}
    \caption{The schematic representation of the KFTD interpolation model (First stage). As an innovative approach, this model takes the initial state $x_{t_0}$ and the final state $x_{t_h}$ as inputs, and generates the intermediate interpolated states $x_{t_j}$.}
    \label{fig2}
\end{figure*}

\subsection{Ocean Spatiotemporal Forecasting}

Artificial intelligence has exhibited considerable potential across numerous domains \cite{1,2,3,4,5,6,7,8,9,10,11,12,13,14,15,16,17}. Paulo et al. \cite{1de2022hybrid} introduced a hybrid system approach for preliminary sea surface temperatures(SST) forecasting using a base model; subsequently, the model's residuals (prediction errors) are modeled to refine and enhance the initial forecasts. Nabila et al. \cite{2zrira2024time} combined BiLSTM with attention mechanisms for accurate SST prediction, but these methods neglected the spatial characteristics. For regional map forecasting with spatial information, Yang et al. \cite{9yang2017cfcc} utilized fully connected LSTM (FC-LSTM) layers and convolutional layers to extract spatiotemporal features.

Zheng et al. \cite{10zheng2020purely} cascaded and stacked convolutional layers to predict the subsequent timestep's spatiotemporal map. Employing an iterative forward prediction method, they analyzed the forecast results for complex tropical instability waves. Shao et al. \cite{11shao2021deep} proposed a forecasting model for multi-marine variable prediction to imbue data-driven methods with physical significance. It integrated Multivariate Empirical Orthogonal Function analysis with Conv1D-LSTM, combining the strengths of one-dimensional convolution operations and LSTM neural networks. However, marine data itself implies complex spatial and temporal processes and dynamic changes of multiple elements. When the data's complexity exceeds the model's ability to capture it, effectiveness will be significantly reduced. Deep learning approaches are purely data-driven \cite{12wang2023data}, relying solely on the information contained within the data and not requiring physical mechanisms or external forces as model inputs \cite{13lutter2023combining}. Consequently, the complexity of the information in the data directly influences model performance. When the complexity of the data surpasses the model's capacity to capture it, the model's effectiveness is compromised \cite{14kim2023multi}.

Our work builds upon these areas by introducing a novel time-differentiable two-stage framework that integrates the Koopman operator and Fourier Analysis Perceptron to capture complex nonlinear dynamics and periodic patterns.

\subsection{Koopman Operator}
The Koopman operator provides a linear representation of nonlinear dynamical systems \cite{19peitz2024equivariance}. By lifting the system into an infinite-dimensional function space, the Koopman operator allows for the analysis of nonlinear systems using linear techniques \cite{20shi2024koopman}. This operator has been applied in various domains, including fluid dynamics and control theory, to predict the evolution of complex systems \cite{22arbabi2017study}. Recently, there has been interest in leveraging the Koopman operator within machine learning models to enhance their ability to capture and predict the dynamics of complex systems \cite{21dogra2020optimizing}. Chen et al. \cite{37chen2024forecasting} discussed an approach to dynamics prediction via incomplete equations of motion and the self-encoder Koopman operator. This approach works by lifting the nonlinear dynamics into Koopman space. Mallen et al. \cite{36mallen2024deep} proposed a technique called Deep Probabilistic Koopman (DPK), which is based on the theory of Koopman operators for implementing long-term time series forecasting. This shows the potential of Koopman's operator theory in dealing with long-term forecasting problems with periodic uncertainty. Although there are no mature cases of direct application to marine spatiotemporal forecasting, the successful applications in the above fields provide a theoretical basis and technical reference. 

Our work leverages the Koopman operator to transform complex nonlinear dynamical problems into infinite-dimensional linear systems, achieving global linearization and aiding the model in understanding complex ocean dynamics.

\section{Koopman-Fourier Time-Differentiable Network}

\subsection{Problem Definition}
In ocean spatiotemporal forecasting, we commonly use time series, where each temporal instance corresponds to a distribution of temperatures across a spatial domain. Formally, the ocean element prediction problem can be articulated as follows: Given a time series $\left\{ x_{t-i} \right\}_{i=1}^{k}$, where $x_{t-i}$ denotes the spatiotemporal distribution at time $t-i$. This encompasses longitude, latitude, and additional features that may correlate. Our goal is to predict the distribution at subsequent time points $t+1, t+2, \ldots, t+h$, and to generate the forecasted outputs $\left\{ \hat{x}_{t+j} \right\}_{j=0}^{h}$, where $\hat{x}_{t+j} \in \mathbb{R}^2$.

\subsection{KFTD Model}
We have abandoned the traditional diffusion model approach due to its high computational cost, which requires sampling through hundreds of diffusion steps during both training and inference. Conversely, we have adopted an interpolation-based prediction process. Our model training is divided into two stages. 

\textbf{In the first stage}, as shown in Algorithms \ref{s1}, we train an interpolation model to simulate the noise addition process. The goal is to generate intermediate states between known time points. the interpolation model takes three inputs: a randomly sampled time step \(j\) from the uniform distribution \(\{1, \ldots, h-1\}\), the initial state \(X_{t_0}\), and the target end state \(X_{t_h}\). The objective is to minimize the loss between the model's output \(X_j\) at time step \(j\) and the true value at that time step. After training, the interpolation model can perform precise interpolations between any two known time points. As shown in Figure \ref{fig2}, The model begins with a convolutional layer for feature extraction. The encoder module segments the data and adds positional encoding to enhance spatial awareness. KoopmanNet enhances the Koopman operator's representational power by extracting temporal periodic features through the FAP. The data then moves to the decoder module, where residual blocks and temporal step embedders refine the data interpretation, preparing the model to generate the interpolated state at the \(j\)-th time step. The final output is produced by a final convolutional layer.

\begin{figure}
    \centering
    \includegraphics[width=1\linewidth]{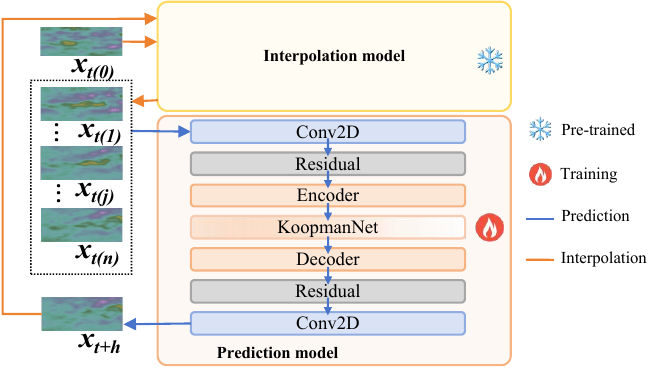}
    \caption{The schematic representation of the KFTD prediction model (Second stage). This model takes the intermediate state $x_{t_j}$ as inputs, and generates the final state $x_{t_h}$}
    \label{fig3}
\end{figure}

\textbf{In the second stage}, as shown in Algorithms \ref{s2} and Figure \ref{fig3}, we train a prediction model to simulate the denoising process. This model uses the interpolation model from the first stage to generate intermediate states and then predicts the target end state. It takes the generated \(j\)-th intermediate state \(X_{t_j}\) as input and outputs the predicted target end state \(X_{t_h}\). The training objective is to minimize the difference between the predicted end state \(X_{t_h}\) and the actual value.

\begin{algorithm}[tb]
\caption{Train interpolator network, \(\mathcal{I}_\phi\)}
\label{s1}
\begin{algorithmic}[1]
\STATE \textbf{Input:} networks \(\mathcal{I}_\phi\), norm \(\|\cdot\|\), horizon \(h\), schedule \([j_n]_{j=0}^{N-1}\)
\STATE Sample \(i \sim \text{Uniform}(\{1, \ldots, h-1\})\)
\STATE Sample \(x_t, x_{t+i}, x_{t+h} \sim \mathcal{X}\)
\STATE Optimize \(\min_{\phi}L_{total}(\mathcal{I}_{\phi}(x_t, x_{t+h}, j), x_{t+j})\)
\STATE \textbf{Output:} Trained networks \(\mathcal{I}_\phi\)
\end{algorithmic}
\end{algorithm}

\begin{algorithm}[tb]
\caption{Train forecaster network, \(F_\theta\)}
\label{s2}
\begin{algorithmic}[1]
\STATE \textbf{Input:} Trained networks \(F_\theta\), networks \(\mathcal{I}_\phi\), norm \(\|\cdot\|\), horizon \(h\), schedule \([j_n]_{j=0}^{N-1}\)
\STATE Freeze \(\mathcal{I}_{\phi}\) and enable inference stochasticity (e.g., dropout)
\STATE Sample \(n \sim \text{Uniform}(\{0, \ldots, N-1\})\) and \(x_t, x_{t+h} \sim \mathcal{X}\)
\STATE Optimize \(\min_{\theta} L_{total}(F_{\theta}(\mathcal{I}_{\phi}(x_t, x_{t+h}, j_n), j_n), x_{t+h})\)
\STATE \textbf{Output:} Trained networks \(F_\theta\)
\end{algorithmic}
\end{algorithm}

The KFTD model innovatively decouples temporal interpolation from spatiotemporal prediction through a two-stage training paradigm that replaces high-cost diffusion sampling with a lightweight, differentiable interpolation mechanism. By leveraging FAP-based KoopmanNet, it achieves continuous-time sub-step generation without fixed grids, enabling adaptive resolution and enforcing physical priors via modular residual terms. This architecture yields a 4× speed-up over diffusion baselines while maintaining physical consistency, breaking the fidelity-efficiency trade-off in operational ocean forecasting.

\subsection{Feature Processing and Encoding}
\label{app:feature_processing}
To enhance the model's understanding of data, input data $t$ and conditions $c$ are concatenated, adding context.

The convolutional layer $C$ extracts features from data, capturing local patterns. Temporal data $t$ is processed with a multi-layer perceptron and then combined with initial features in a residual module.

The PatchEmbed module divides data into patches and maps them into a high-dimensional space, 
$x' = \mathcal{P}a(x)$, where $x$ is the input, $x'$ is the output, and $\mathcal{P}a$ is the PatchEmbed module. It handles spatiotemporal data, dividing it into patches and mapping them into a high-dimensional space. The input data $x$ is a four-dimensional tensor with dimensions $(\text{Batch}, \text{Channels}, \text{Lon}, \text{Lat})$. PatchEmbed divides it into $P_{Lon} \times P_{Lat}$ patches, each of size $(p_{Lon}, p_{Lat})$, where $x'$ has dimensions $(B, N, E)$.

The patch embedding is calculated as follows:
\begin{equation}
x'_{b,n,e} = \sum_{i=0}^{p_{Lon}-1} \sum_{j=0}^{p_{Lat}-1} x_{b,c,\text{lat}(n)+i,\text{Lat}(n)+j} \cdot w_e
\end{equation}
$x'_{b,n,e}$ is the embedded feature value, $\text{lat}(n)$ and $\text{Lat}(n)$ are indices, and $w_e$ is the embedding weight.

Positional encoding adds spatial information to the embedded representation, capturing spatial structure,
$\mathcal{P} = (\mathcal{P}_1, \mathcal{P}_2, \ldots, \mathcal{P}_n)$, 
$\mathcal{P}_n \in \mathbb{R}^E$ is the positional vector for patch $n$. The feature representation with positional information is $x'' = x' + \mathcal{P}$, the positional encoding vector $\mathcal{P}$ is learned during training and can be initialized with sine and cosine values, helping the model understand the data's spatial structure.
\subsection{KoopmanNet}

We design the Koopman Adaptive Fourier Neural Operator (KAFNO) to construct a linear embedding space for nonlinear systems and innovate the Fourier Analysis Perceptron (FAP) to extract multi-scale periodic features, achieving efficient interpolation at any time step. To lower the high computational cost of traditional diffusion models, we propose a hierarchical feature fusion mechanism in each FAP layer, creating a positive correlation between network depth and Fourier coefficient learning capacity.

\subsubsection{Koopman Adaptive Fourier Neural Operator (KAFNO)}

The classical Koopman operator theory lifts nonlinear dynamics into an infinite-dimensional linear system by defining a linear operator acting on observable functions. However, in strongly nonlinear systems, the global linearization assumption often fails due to the lack of a globally invariant linear subspace, leading to exponential error accumulation over time.

To address this, we decompose the Koopman operator into a \emph{multi-scale linear superposition}, inspired by recent extensions of Dynamic Mode Decomposition (DMD) and Hankel-based Koopman methods~\cite{brunton2017chaos,22arbabi2017study}:
\begin{equation}
\mathcal{K} = \mathcal{K}_1 \oplus \mathcal{K}_2 \oplus \dots \oplus \mathcal{K}_L
\end{equation}
Each $\mathcal{K}_l$ corresponds to a distinct temporal scale (e.g., fast-varying vs. slow-varying modes), and the modal partitioning is learned automatically via neural networks. This decomposition is theoretically grounded in the \emph{Hankel Alternative View of Koopman (HAVOK)} framework~\cite{brunton2017chaos}, which shows that complex nonlinear dynamics can be approximated by a linear model plus a low-rank nonlinear forcing term in delay-embedded space.

Formally, we define learnable projection matrices $W_l \in \mathbb{R}^{d_l \times d}$ that map the state $x$ into the $l$-th subspace:
\begin{equation}
z_l = W_l x, \quad \mathcal{K}_l z_l = A_l z_l \quad (A_l \in \mathbb{R}^{d_l \times d_l})
\end{equation}
This construction enables the model to capture both global linear dynamics and localized nonlinear perturbations, extending the applicability of Koopman theory to strongly nonlinear systems.

\subsubsection{Fourier Analysis Perceptron}

In order to extract periodic characteristics, we designed a Fourier analysis prior (FAP) to process $y$. FAP functions into constituent frequencies, thereby revealing underlying periodic structures. The Fourier series represents a periodic function as an infinite sum of sines and cosines:
\begin{equation}
f_s(y) = a_0 + \sum_{n=1}^{N} \left( a_n \cos\left(\frac{2\pi ny}{T}\right) + b_n \sin\left(\frac{2\pi ny}{T}\right) \right).
\end{equation}

A simple Fourier series neural network can be expressed as:
\begin{equation}
f_s(y) = B + \mathbf{W}_{\text{out}} \left[ \cos(\mathbf{W}_{\text{in}} y) \Vert \sin(\mathbf{W}_{\text{in}} y) \right],
\end{equation}
where \( B \in \mathbb{R}^{d_y} \), \( \mathbf{W}_{\text{in}} \in \mathbb{R}^{N \times d_x} \), and \( \mathbf{W}_{\text{out}} \in \mathbb{R}^{d_y \times 2N} \) are learnable parameters, and \( \Vert \) denotes the concatenation along the first and second dimension. When multiple \( f_s \) are stacked, they form a deep neural network: \( f_D(y) = f_s(f_s(\ldots f_s(y) \ldots)) \). 

However, we have observed that the direct stacking of \( f_s(y) \) leads the primary parameters of the model \( f_D(x) \) to focus on learning the angular frequencies (\( \omega_n = \frac{2\pi n}{T} \)), thereby neglecting the learning of the Fourier coefficients (\( a_n \) and \( b_n \)). This issue is further exacerbated in deep networks \( f_D(x) = l_L \circ \cdots \circ l_1(x) \), where parameters \( W_{\text{in}}^i \) and \( W_{\text{out}}^i \) progressively bias toward angular frequency learning, while the Fourier coefficient learning capacity shows no dependence on the depth of the network. For example, the output of the \( L \)-th layer becomes:
\begin{equation}
f_D(x) = B^L + W_{\text{out}}^L \left[ \cos(W_{\text{in}}^L (l_{1:L-1} \circ x)) \| \sin(W_{\text{in}}^L (l_{1:L-1} \circ x)) \right].
\end{equation}
Where \( W_{\text{in}}^L \) dominates angular frequency learning, while \( W_{\text{out}}^L \) merely serves as a final-layer linear combination, failing to enhance coefficient expressivity through depth.

To address these limitations, we designed the FAP to ensure that the ability to learn Fourier coefficients is positively correlated with the depth of the neural network, while allowing the outputs of intermediate layers to be modeled periodically by Fourier series and maintaining the capacity for nonlinearity in slightly periodic systems. The FAP is defined as:
\begin{equation}
\begin{aligned}
\phi(y) &\triangleq \left[ \cos(\mathbf{W}_p y) \Vert \sin(\mathbf{W}_p y) \right] \sigma(\mathbf{B}_{\bar{p}} + \mathbf{W}_p y) \\
&+ \sigma(\mathbf{B}_m + \mathbf{W}_m y),
\end{aligned}
\end{equation}
where \( \sigma \) denotes an activation function. This approach ensures the correlation of Fourier coefficients with depth, thereby enhancing the ability to extract periodic information. Additionally, \( \sigma(\mathbf{B}_m + \mathbf{W}_m y) \) prevents overfitting to periodicity, ensuring the extraction of both periodic and aperiodic information in complex ocean spatiotemporal forecasting.

In KoopmanNet, the FAP processes the output \( y \) after the inverse Fourier transform. The function \( f \) is defined as:
\begin{equation}
f_\phi = \phi(f_{\mathcal{F}^{-1}}),
\end{equation}
where \( \phi \) is the FAP operation. Furthermore, a high-dimensional vector \( f_k \) is computed as \( f_k = f_\phi + f_{cov} \).

Two crucial innovations are the explicit correlation between Fourier coefficients and network depth and the periodic-aperiodic separation mechanisms. We design a deep network to establish a positive correlation between Fourier coefficient learning capacity and network depth while preserving universal modeling capabilities.

In KoopmanNet, the FAP processes the output after the inverse Fourier transform. The function $f$ is defined as $f_\phi = \phi(f_{\mathcal{F}^{-1}})$, where $\phi$ is the FAP operation. A high-dimensional vector $f_k$ is computed as $f_k = f_\phi + f_{cov}$. The FAP achieves superior frequency modulation and coefficient relationships compared to shallow Fourier networks.

To achieve depth-enhanced Fourier coefficient learning, we employ hierarchical feature fusion in each FAP layer \( \phi_l(x) \):
\begin{equation}
\scriptstyle
\phi_l(x) = \left[ \cos(W_p^l x) \| \sin(W_p^l x) \right] \odot \sigma(B_{\bar{p}}^l + W_p^l x) + \sigma(B_m^l + W_m^l x),
\end{equation}
where \( \odot \) denotes element-wise multiplication, and \( \sigma \) denotes the activation function. The terms \( \cos(W_p^l x) \) and \( \sin(W_p^l x) \) explicitly model frequencies. Adjustments to periodic signal amplitudes are made via \( \sigma(B_{\bar{p}}^l + W_p^l x) \), which is equivalent to learning \( a_n \) and \( b_n \). The term \( \sigma(B_m^l + W_m^l x) \) preserves universal modeling capacity, while residual connections prevent over-reliance on periodicity assumptions.

The input to the \( (l+1) \)-th layer becomes:$x_{l+1} = \phi_l(x_l)$

Subsequent layers dynamically enhance or suppress specific frequency components through \( W_p^{l+1} \) while learning complex Fourier coefficient combinations. Hierarchical stacking of \( \sigma \) functions achieves depth accumulation of Fourier coefficient amplitudes.

For an input \( x \in \mathbb{R} \), the first layer outputs:

\begin{align}
\phi_1(x) &= \left[ \cos(w_{p1} x) \cdot \sigma(w_{p1} x + b_1) \right. \nonumber \\
&\qquad \left. \| \sin(w_{p1} x) \cdot \sigma(w_{p1} x + b_1) \right] \nonumber \\
&\quad + \sigma(w_{m1} x + b_{m1}).
\end{align}

The second layer processes these features as:

\begin{align}
\phi_2(\phi_1)
&= \Bigl[ \cos(W_p^{2}\phi_1)\odot\sigma(W_p^{2}\phi_1+B_{\bar p}^{2}) \nonumber\\
&\qquad \bigl\| \sin(W_p^{2}\phi_1)\odot\sigma(W_p^{2}\phi_1+B_{\bar p}^{2}) \Bigr] \nonumber\\
&\quad + \sigma(W_m^{2}\phi_1+B_m^{2}).
\end{align}

Where \( W_p^2 \) learns combined frequencies (e.g., \( \cos(\omega_1 x \pm \omega_2 x) \)) from the periodic signals of \( \phi_1 \), while \( \sigma \) functions enable hierarchical amplitude modulation. Through multilayer stacking, each layer's output contains both periodic and aperiodic features, allowing subsequent layers to dynamically adjust their proportion. The FAP achieves superior frequency modulation and coefficient relationships compared to shallow Fourier networks.

The decoding procedure translates the feature maps processed by KoopmanNet back into the original data space. Initially, the feature map \( k \) is subjected to a transposed convolutional layer, which adjusts its spatial dimensions to align with those of the initial input data. Subsequently, the predicted target state dimensions are generated through a residual connection and a final convolutional operation.

\subsection{Data-physics Prior Loss}

The proposed Data–Physics Prior loss pioneers a modular, residual-based regularization paradigm that seamlessly injects arbitrary differentiable physical laws into end-to-end training, transcending the conventional penalty-weight balancing act to yield a plug-and-play conservation constraint engine with provable Sobolev regularity. 

This innovative approach not only ensures that the model learns the statistical properties of the data but also incorporates underlying physical mechanisms, thereby enhancing its generalizability and reliability. The D-PP loss function combines a physics-based loss component with a data-driven loss component, allowing for the training of interpolation and prediction networks that are both accurate and physically consistent.

To balance the data loss and the physics-informed loss, a hyperparameter $\lambda$ is introduced, allowing us to adjust the weights of both in the total loss function. The total loss is then expressed as:

\begin{equation}
\mathcal{L}_{total} = \lambda \mathcal{L}_{data} + (1 - \lambda) \mathcal{L}_{physics}
\end{equation}
This approach guarantees that model predictions are accurate and align with physical principles, resulting in robust and reliable outcomes, especially in data-scarce or noisy conditions. The hyperparameter \( \lambda \) fine-tunes the impact of physical laws in training, beneficial for applications requiring high physical precision.

To move beyond L2-based constraints, we incorporate Sobolev regularity:

\begin{equation}
\mathcal{L}_{\text{Sobolev}} = \frac{1}{M} \sum_{i=1}^{M} \left( \left\| \nabla \hat{u}_i \right\|_{L^2}^2 + \left\| \nabla^2 \hat{u}_i \right\|_{L^2}^2 \right).
\end{equation}

The design objective of the D-PP loss function is to support arbitrary differentiable physical equations. To achieve this, we generalize the loss formulation as:

\begin{equation}
\mathcal{L}_{\text{D-PP}} = \frac{1}{M} \sum_{i=1}^{M} \left( \mathcal{R}\left( \hat{u}_i, \theta \right) \right)^2,
\end{equation}

where \( \mathcal{R} \) represents the residual of the physical equation parameterized by \( \theta \). This formulation allows the loss function to adapt to different physical models by simply changing the residual function \( \mathcal{R} \). 

Here is an example: The physical loss based on the heat conduction equation, is calculated using the finite difference method to approximate the second-order spatial derivative of the input tensor u on the specified dimension. This regularization term prevents overfitting to training data noise, thereby improving model stability. For one-dimensional space,
\begin{equation}
u_{ss} = \frac{\hat{u}(s+h) - 2\hat{u}(s) + \hat{u}(s-h)}{h^2}
\end{equation}
Here, $h$ is the spatial step size. The first time derivative is approximated similarly when time-dependent data is available:
\begin{equation}
     u_\tau = \frac{\hat{u}(\tau+\Delta \tau) - \hat{u}(\tau)}{\Delta \tau} 
\end{equation}
where $\Delta$ is the time step, the physical prior loss is a residual based on the laws of physics, where we consider the heat conduction equation. We calculate the temperature distribution predicted by the model and compare it with the analytical solution of the heat conduction equation.

\begin{equation}
\mathcal{L}_{\text{physics}} = \frac{1}{M} \sum_{i=1}^{M} (u_\tau-\alpha u_{ss} )^2
\end{equation}
where $\alpha$ is the thermal diffusivity.

Data loss is calculated by comparing the model's predicted output $\hat{y}$ with the true target values $y$. We use the mean squared error as the measure of data loss, expressed as:

\begin{equation}
\mathcal{L}_{data} = \frac{1}{M} \sum_{i=1}^{M} (z_{i} - \hat{z}_{i})^2
\end{equation}

where $N$ is the number of samples, $y_i$ is the true value of the $i^{th}$ sample, and $\hat{z}_{i}$ is the predicted value of the $i^{th}$ sample by the model.

\begin{table*}
\centering

\caption{Comparison of the performance of different methods in forecasting different oceanic elements. \textbf{Bold} values indicate the best performance. \textcolor{blue}{Blue} values indicate the second best.}
\label{table:performance_comparison}
\resizebox{1.\textwidth}{!}{
\begin{tabular}{lcccccccccccc}
\toprule
\textbf{Method} & \multicolumn{3}{c}{\textbf{OISSTv2}} & \multicolumn{3}{c}{\textbf{SWPT10M}} & \multicolumn{3}{c}{\textbf{SWPT50M}} & \multicolumn{3}{c}{\textbf{SSHG}} \\
\cmidrule(lr){2-4} \cmidrule(lr){5-7} \cmidrule(lr){8-10} \cmidrule(lr){11-13}
& \textbf{CRPS (std)} & \textbf{MAE (std)} & \textbf{MSE (std)} & \textbf{CRPS (std)} & \textbf{MAE (std)} & \textbf{MSE (std)} & \textbf{CRPS (std)} & \textbf{MAE (std)} & \textbf{MSE (std)} & \textbf{CRPS (std)} & \textbf{MAE (std)} & \textbf{MSE (std)} \\
\midrule
DDPM \cite{23ho2020denoising} & 0.342 (0.025) & 0.184 (0.012) & 0.214 (0.018) & 0.425 (0.031) & 0.324 (0.021) & 0.357 (0.029) & 0.437 (0.033) & 0.274 (0.019) & 0.265 (0.022) & 0.0196 (0.0002) & 0.00064 (0.000014) & 0.00077 (0.000015) \\
MCVD \cite{24voleti2022mcvd} & \textcolor{blue}{0.299 (0.021)} & 0.162 (0.010) & 0.175 (0.015) & 0.388 (0.028) & 0.309 (0.019) & 0.293 (0.023) & 0.374 (0.026) & 0.252 (0.017) & 0.230 (0.019) & 0.0215 (0.0003) & 0.00068 (0.000013) & 0.00080 (0.000014) \\
Dyffusion \cite{18ruhling2024dyffusion} & 0.318 (0.023) & 0.167 (0.011) & 0.186 (0.016) & 0.351 (0.025) & 0.288 (0.018) & 0.283 (0.021) & 0.405 (0.029) & 0.243 (0.016) & 0.236 (0.018) & 0.0203 (0.0002) & \textcolor{blue}{0.00055 (0.000012)} & 0.00075 (0.000013) \\
STEMO \cite{48shao2024stemo} & 0.310 (0.020) & 0.150 (0.009) & 0.190 (0.016) & 0.365 (0.025) & 0.260 (0.015) & 0.285 (0.020) & 0.380 (0.028) & 0.235 (0.014) & 0.250 (0.017) & 0.0190 (0.0002) & 0.00060 (0.000013) & 0.00070 (0.000013) \\
UniST \cite{47yuan2024unist} & 0.295 (0.018) & 0.140 (0.007) & 0.178 (0.013) & 0.350 (0.023) & 0.245 (0.013) & 0.270 (0.019) & 0.370 (0.025) & 0.225 (0.012) & 0.240 (0.015) & 0.0175 (0.0001) & 0.00055 (0.000013) & 0.00065 (0.000012) \\
VMRNN \cite{25tang2024vmrnn} & 0.324 (0.024) & 0.154 (0.009) & 0.183 (0.017) & \textcolor{blue}{0.337 (0.022)} & 0.254 (0.015) & 0.271 (0.020) & \textcolor{blue}{0.363 (0.024)} & \textcolor{blue}{0.217 (0.013)} & \textbf{0.221 (0.016)} & 0.0184 (0.0002) & 0.00065 (0.000011) & 0.00079 (0.000013) \\
RPMixer \cite{26yeh2024rpmixer} & 0.303 (0.022) & \textcolor{blue}{0.136 (0.008)} & \textcolor{blue}{0.173 (0.014)} & 0.340 (0.026) & \textcolor{blue}{0.237 (0.014)} & \textcolor{blue}{0.247 (0.017)} & 0.372 (0.027) & 0.231 (0.015) & 0.246 (0.018) & \textcolor{blue}{0.0188 (0.0002)} & 0.00064 (0.000012) & \textcolor{blue}{0.00058 (0.000012)} \\
Ours & \textbf{0.287 (0.019)} & \textbf{0.121 (0.007)} & \textbf{0.151 (0.012)} & \textbf{0.325 (0.020)} & \textbf{0.231 (0.013)} & \textbf{0.245 (0.015)} & \textbf{0.356 (0.023)} & \textbf{0.214 (0.012)} & \textcolor{blue}{0.227 (0.014)} & \textbf{0.0164 (0.0001)} & \textbf{0.00054 (0.000012)} & \textbf{0.00061 (0.000011)} \\

\bottomrule
    \end{tabular}}
\end{table*}

\section{Experiments}
We first describe our experimental setup, perform ablation studies, and compare our method with other spatiotemporal forecasting methods. We then present the visualization results of oceanic spatiotemporal forecasting and discuss the findings. The code will be released upon acceptance.

\subsection{Experimental Setup}
To ensure fair comparisons, we defined the experimental task as inputting a single initial state $t_0$ and predicting the subsequent 7-step states, meaning the model will generate seven future states. We implemented KFTD within the PyTorch framework and trained it using the Adam optimizer with a learning rate 0.0001 and a batch size 32. All experiments, including the other deep learning methods used for comparisons, were performed on NVIDIA RTX A6000 GPUs. The seed for training and testing was fixed, and the $\lambda$ value in the loss function was set to 0.6. To ensure the credibility of the model's test results, we pooled and averaged the results of the 20 tests. Additionally, we calculated the standard deviations of the performance metrics and performed paired t-tests to assess the statistical significance of the performance differences between our method and the best baseline. All reported improvements of our method over the best baseline are statistically significant (p < 0.05) unless otherwise noted.

\subsubsection{Datasets}
We build multi-regional benchmarks from NOAA OISSTv2 and Copernicus GLORYS12 reanalyses to test generalizability across four distinct oceanic regimes. Each regime is partitioned into 60 × 60 grid tiles at the native resolution, and we randomly sample several tiles per regime:
(A) Tropical Pacific (10° S–10° N, 170–110° W);
(B) North Atlantic (35–55° N, 30–60° W);
(C) Equatorial Indian Ocean (10° S–10° N, 50–80° E);
(D) Southern Ocean (60–75° S, 20–90° E).

\begin{itemize}[leftmargin=*,nosep]
\item \textbf{OISSTv2}: daily NOAA 0.25° SST, 1982–2019 train, 2020 val, 2021 test.
\item \textbf{SWPT10M}: GLORYS12 0.083° 10 m potential temperature, 1998–2020 train, 2021 val, 2022-H1 test.
\item \textbf{SWPT50M}: identical to SWPT10M but at 50 m depth.
\item \textbf{SSHG}: GLORYS12 sea-surface height above geoid, same temporal split.
\end{itemize}

\subsubsection{Baselines}

We compared our method with several state-of-the-art spatiotemporal forecasting models, including DDPM \cite{23ho2020denoising}, MCVD \cite{24voleti2022mcvd}, DYffusion \cite{18ruhling2024dyffusion}, VMRNN \cite{25tang2024vmrnn}, RPMixer \cite{26yeh2024rpmixer}, STEMO \cite{48shao2024stemo}, and UniST \cite{47yuan2024unist}.



\subsection{Comparison with State-Of-The-Art Methods}

In our comparative analysis, we evaluated our approach against existing spatiotemporal forecasting methods and other diffusion models, as shown in Table \ref{table:performance_comparison}. Our model’s improvements over the best-performing baseline are statistically significant across all datasets (p < 0.01).

Notably, our method outperforms existing spatiotemporal forecasting models in all evaluation metrics. This result not only highlights the potential of diffusion models in spatiotemporal forecasting but also confirms the effectiveness of the Koopman operator in improving the prediction accuracy of diffusion models. This effectiveness is attributed to its ability to transform nonlinear problems into linear ones, thus simplifying the model training and prediction process. Our model effectively captures local details and overall trends in spatiotemporal data through multiscale analysis techniques. This multiscale property makes the model more effective in dealing with data with complex spatiotemporal dependencies. Experiments on various scenarios and datasets demonstrate our model has good generalization ability. 

\begin{figure*}
    \centering
    \includegraphics[width=0.9\linewidth]{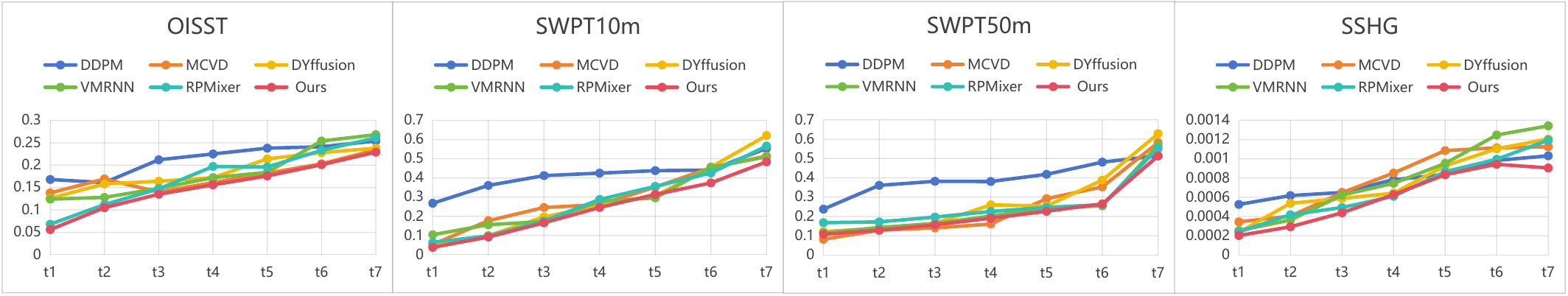}
    \caption{MSE analysis of predictive models over a seven-step horizon across four datasets. Our model consistently exhibits lower MSE, underlining its enhanced predictive accuracy and reliability in handling complex spatiotemporal dependencies.}
    \label{fig5}
\end{figure*}

Figure \ref{fig5} illustrate the MSE curves for comparative models over a predictive horizon of one to seven steps. The data show distinct MSE profiles, with a general trend of increasing MSE as the number of prediction steps grows, indicating rising predictive uncertainty or error.

In all datasets, our model (labeled \textit{Ours}) consistently exhibits lower MSE values across most predictive steps, outperforming others. \textit{Ours} and \textit{RPMixer} maintain a stable MSE increases, indicating reliability. In contrast, \textit{Dyffusion} has a steeper MSE rise in the SWPT50M dataset, suggesting instability. Traditional diffusion models like DDPM generally underperform, with higher error margins in some scenarios, showing a poor fit for the tasks.

\subsection{Long-Term Forecasting Performance}
We evaluate the ability of KFTD and competitive baselines to generate continuous 6-month forecasts on the OISSTv2 dataset. The task is initialized with a single monthly snapshot and predicts 6 successive months. Results are reported in Table~\ref{tab:longterm_6m}. KFTD maintains the lowest error growth across all lead times, confirming its advantage for seasonal-scale ocean prediction.
\begin{table*}[ht]
\centering

\caption{Six-month forecasting performance on OISSTv2. Last column shows the 6-month average.}
\label{tab:longterm_6m}
\resizebox{0.9\textwidth}{!}{
\begin{tabular}{lcccccccccccccc}
\toprule
\textbf{Method} & \multicolumn{2}{c}{\textbf{Month 1}} & \multicolumn{2}{c}{\textbf{Month 2}} & \multicolumn{2}{c}{\textbf{Month 3}} & \multicolumn{2}{c}{\textbf{Month 4}} & \multicolumn{2}{c}{\textbf{Month 5}} & \multicolumn{2}{c}{\textbf{Month 6}} & \multicolumn{2}{c}{\textbf{6-Month Avg}} \\
\cmidrule(lr){2-3}\cmidrule(lr){4-5}\cmidrule(lr){6-7}\cmidrule(lr){8-9}\cmidrule(lr){10-11}\cmidrule(lr){12-13}\cmidrule(lr){14-15}
 & CRPS & MSE & CRPS & MSE & CRPS & MSE & CRPS & MSE & CRPS & MSE & CRPS & MSE & CRPS & MSE \\
\midrule
DDPM & 0.418 & 0.286 & 0.472 & 0.331 & 0.527 & 0.383 & 0.583 & 0.442 & 0.638 & 0.497 & 0.694 & 0.554 & 0.555 & 0.416 \\
MCVD & 0.375 & 0.241 & 0.421 & 0.284 & 0.472 & 0.329 & 0.523 & 0.376 & 0.575 & 0.422 & 0.628 & 0.469 & 0.499 & 0.353 \\
DYffusion & 0.392 & 0.255 & 0.440 & 0.299 & 0.493 & 0.344 & 0.546 & 0.391 & 0.599 & 0.438 & 0.653 & 0.485 & 0.521 & 0.369 \\
STEMO & 0.385 & 0.261 & 0.432 & 0.305 & 0.485 & 0.349 & 0.539 & 0.396 & 0.592 & 0.443 & 0.646 & 0.490 & 0.513 & 0.374 \\
UniST & 0.370 & 0.245 & 0.415 & 0.287 & 0.467 & 0.332 & 0.520 & 0.378 & 0.573 & 0.425 & 0.626 & 0.472 & 0.495 & 0.356 \\
VMRNN & 0.399 & 0.250 & 0.447 & 0.294 & 0.500 & 0.338 & 0.553 & 0.385 & 0.606 & 0.431 & 0.659 & 0.478 & 0.527 & 0.363 \\
RPMixer & 0.381 & 0.239 & 0.426 & 0.281 & 0.478 & 0.325 & 0.531 & 0.372 & 0.584 & 0.418 & 0.637 & 0.465 & 0.506 & 0.350 \\

\textbf{Ours} & \textbf{0.360} & \textbf{0.220} & \textbf{0.401} & \textbf{0.258} & \textbf{0.448} & \textbf{0.299} & \textbf{0.497} & \textbf{0.341} & \textbf{0.546} & \textbf{0.384} & \textbf{0.595} & \textbf{0.428} & \textbf{0.474} & \textbf{0.322} \\
\bottomrule
\end{tabular}}
\end{table*}

To better contextualize long-term forecasting performance, we introduce a climatological mean baseline (averaging the same calendar months over the previous 5 years). Table \ref{tab:climatology} reports MSE results for this baseline and KFTD over 2/4/6-month horizons.
\begin{table}[ht]
\centering

\caption{Long-term forecasting performance comparison with climatological mean baseline.}
\label{tab:climatology}
\resizebox{1.0\columnwidth}{!}{
\begin{tabular}{lcccc}
\toprule
\textbf{Method} & \textbf{2-Month MSE} & \textbf{4-Month MSE} & \textbf{6-Month MSE} & \textbf{Avg MSE} \\
\midrule
Climatology & 0.783 & 0.791 & 0.715 & 0.762 \\
Ours (KFTD) & 0.258 & 0.341 & 0.428 & 0.322 \\
\bottomrule
\end{tabular}}
\end{table}

KFTD substantially reduces long-term forecasting errors compared with the climatological baseline, confirming its ability to capture interannual anomalies and dynamic deviations beyond seasonal climatology. Notably, the climatology baseline does not exhibit increasing error with forecast lead time (as it predicts the stationary seasonal mean), while KFTD performs true temporal extrapolation—generating evolving ocean states consistent with physical dynamics.

To further substantiate the computational efficiency of KFTD, we report single-batch inference latency and resource consumption metrics (measured on identical NVIDIA RTX A6000 hardware). The single-batch inference latency of MCVD is 79.167 s, while KFTD achieves 18.8021 s (representing a ~4.21× speedup, +76.25\% efficiency gain).

\subsection{Ablation Studies}

To conduct an ablation study, we tested the proposed components on the OISSTv2 dataset, focusing on the model's ability to predict the marine state for the next three days.
\begin{table*}[htbp]
\centering
\caption{Ablation Study of Components on Performance. The $\times$ represents not used, and $\checkmark$ represents used.}
\label{table:ablation_study}
\resizebox{1\textwidth}{!}{
\begin{tabular}{cccccccccccccccccccc}
\toprule
\multicolumn{4}{c}{Components} & \multicolumn{2}{c}{$t_1$} & \multicolumn{2}{c}{$t_2$} & \multicolumn{2}{c}{$t_3$} & \multicolumn{2}{c}{$t_4$} &\multicolumn{2}{c}{$t_5$} &  \multicolumn{2}{c}{$t_6$} &\multicolumn{2}{c}{$t_7$} &  \multicolumn{2}{c}{Avg} \\ \cmidrule(lr){1-4} \cmidrule(lr){5-6} \cmidrule(lr){7-8} \cmidrule(lr){9-10} \cmidrule(lr){11-12}\cmidrule(lr){13-14}\cmidrule(lr){15-16}\cmidrule(lr){17-18}\cmidrule(lr){19-20}
KAFNO & FAP & D-PP loss& $f_{cov}$ & CRPS & MSE & CRPS & MSE & CRPS & MSE & CRPS & MSE& CRPS & MSE& CRPS & MSE& CRPS & MSE& CRPS & MSE \\ \cmidrule(lr){1-4} \cmidrule(lr){5-6} \cmidrule(lr){7-8} \cmidrule(lr){9-10} \cmidrule(lr){11-12}\cmidrule(lr){13-14}\cmidrule(lr){15-16}\cmidrule(lr){17-18}\cmidrule(lr){19-20}
$\times$ & $\times $& $\times$ & \checkmark &0.294 & 0.101 & 0.415 & 0.214 & 0.491 & 0.278 & 0.558 & 0.309 & 0.611 & 0.342 & 0.653 & 0.382 & 0.686 & 0.419 & 0.583 & 0.292 \\
$\checkmark$ & $\times$ & $\times$ &\checkmark &0.197 & 0.067 & 0.276 & 0.133 & 0.329 & 0.188 & 0.381 & 0.217 & 0.415 & 0.232 & 0.443 & 0.252 & 0.468 & 0.273 & 0.358 & 0.195 \\
$\times$ & $\checkmark$ & $\times$ &\checkmark &0.215 & 0.075 & 0.297 & 0.145 & 0.366 & 0.212 & 0.426 & 0.243 & 0.471 & 0.267 & 0.515 & 0.292 & 0.553 & 0.317 & 0.406 & 0.222 \\ 
$\checkmark$ & $\checkmark$ &$\times$ &\checkmark &0.183 & 0.062 & 0.255 & 0.122 & 0.302 & 0.157 & 0.353 & 0.193 & 0.388 & 0.212 & 0.421 & 0.232 & 0.451 & 0.252 & 0.336 & 0.176  \\ 
$\times$ & $\times$ & $\checkmark$ &\checkmark &0.251 & 0.082 & 0.323 & 0.206 & 0.314 & 0.197 & 0.372 & 0.231 & 0.411 & 0.253 & 0.443 & 0.275 & 0.472 & 0.295 & 0.369 & 0.220 \\ 
$\times$ & $\checkmark$ & $\checkmark$ &\checkmark &0.202 & 0.061 & 0.283 & 0.197 & 0.292 & 0.173 & 0.341 & 0.208 & 0.378 & 0.224 & 0.410 & 0.242 & 0.440 & 0.262 & 0.335 & 0.195 \\
$\checkmark$ & $\checkmark$ & $\checkmark$ & \checkmark&\textbf{0.147} & \textbf{0.045} & \textbf{0.215} & \textbf{0.099} & \textbf{0.254} & \textbf{0.138} & \textbf{0.298} & \textbf{0.164} & \textbf{0.333} & \textbf{0.185} & \textbf{0.365} & \textbf{0.204} & \textbf{0.394} & \textbf{0.223} & \textbf{0.287} & \textbf{0.151}  \\
$\checkmark$ & $\checkmark$ & $\checkmark$ &$\times$ &0.197 & 0.066 & 0.241 & 0.152 & 0.281 & 0.161 & 0.325 & 0.187 & 0.361 & 0.203 & 0.395 & 0.219 & 0.427 & 0.235 & 0.318 & 0.175 \\

\bottomrule
\end{tabular}}
\end{table*}

\subsubsection{Koopman Adaptive Fourier Neural Operator}

In this study, KAFNO plays a pivotal role, with its core mechanism involving the computation of the Koopman operator. To gain a deeper understanding of the impact of KAFNO, we designed a series of ablation experiments. In these experiments, we denote by $\times$ the scenarios in which only the encoder and decoder were utilized without the computation of the Koopman operator. The symbol $\checkmark$ indicates the use of the koopmanNet layer. The experimental results, as shown in Table \ref{table:ablation_study}, reveal a significant phenomenon in the second and third rows of the data: even without the use of FAP and D-PP loss, KAFNO can still significantly improve the accuracy of marine spatiotemporal forecasting. This finding robustly demonstrates the critical role of the Koopman operator in ocean spatiotemporal forecasting, effectively enhancing the model's capability to capture complex dynamic features.

\subsubsection{Fourier Analysis Perceptron}

In further ablation studies, we investigate the impact of the FAP module on model performance. The symbol $\checkmark$ denotes the integration of the FAP module within the model, while $\times$ signifies the use of only the MLP layer (Multilayer Perceptron), thus validating the pivotal role of FAP in the forecasting process. By comparing the second and fourth rows of data in Table \ref{table:ablation_study}, we observed a significant effect of FAP throughout the diffusion process of the model.

The introduction of the FAP module, through its periodic transformation mechanism, effectively captures the periodic characteristics of marine spatiotemporal data. Moreover, the depth-related strategies of the FAP module further augment the model's ability to capture spatiotemporal features, contributing significantly to the improvement of predictive accuracy and robustness. The experimental results indicate that the incorporation of the FAP module not only enhances the model's predictive accuracy, but also strengthens the model's sensitivity and adaptability to changes in marine spatiotemporal data.

In the fifth row of Table \ref{table:ablation_study}, we demonstrate the significant enhancement in model predictive accuracy achieved by applying both KAFNO and FAP. This result further confirms the complementarity and synergistic effect of the two components in enhancing model performance. KAFNO optimizes the dynamic representation of features by introducing the Koopman operator. At the same time, FAP strengthens the model's capability to capture spatiotemporal features through the periodic transformation of Fourier series and depth-related strategies. Combining the two improves the model's predictive precision for marine changes and enhances the model's robustness against outliers and noise.

\subsubsection{Data-physics prior loss function}

We analyze the effect when applying KAFNO and FAP simultaneously using data loss and D-PP loss. The data in the fifth rows of Table \ref{table:ablation_study} show an important trend: in terms of prediction accuracy, the data loss scheme that integrates physical a priori knowledge significantly outperforms the scheme that uses only data loss.

Comparing the application effects of these two loss functions, the D-PP loss function incorporating physical prior knowledge can more effectively guide the model in learning the underlying structures and patterns in the data. This strategy improves the model's prediction accuracy of ocean spatial and temporal variations and enhances the model's ability to understand complex ocean dynamic processes. By introducing physical constraints, D-PP loss enables the model to better capture the physical characteristics of ocean spatial and temporal variations during the training process, thus more accurately reflecting the variations in the prediction process. Integrating physical prior knowledge into the data loss function can significantly improve the model's predictive performance. Combining physical knowledge with data-driven methods provides an effective way to improve the accuracy of ocean spatiotemporal forecasting models.

\subsubsection{High-frequency filtered}
In Table \ref{table:ablation_study}, $f_{\text{cov}}$ denotes the use of convolutional layers to retain high-frequency fluctuations. Results show that retaining such information improves model performance, as evidenced by better evaluation metrics when $f_{\text{cov}}$ is true. This suggests that high-frequency information is crucial for capturing complex patterns in ocean spatiotemporal data, enhancing prediction accuracy and reliability.
\begin{figure}
    \centering
    \includegraphics[width=1.1\linewidth]{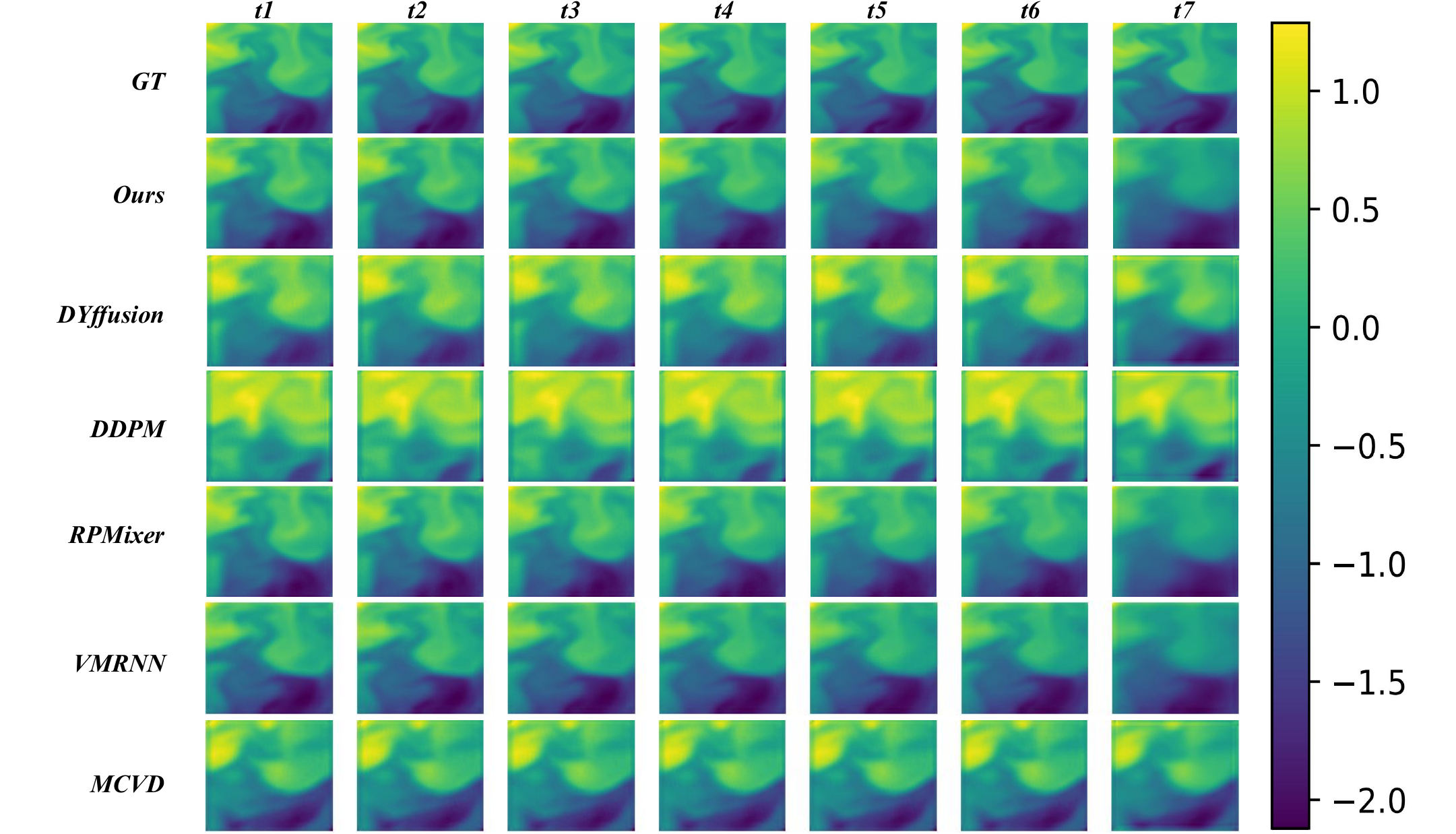}
    \caption{Comparative visualization of the results from spatiotemporal forecasting methods (the 7 time steps for the SWPT50M dataset.}
    \label{fig4}
\end{figure}

\label{otherresults}

\subsection{Qualitative Evaluation}
We visually compared our predictive results with other spatiotemporal prediction methods. Figure \ref{fig4} shows a high degree of congruence between our model's predictions and the actual Ground Truth (GT). Our model accurately captures marine data's macroscopic trends and subtle details, outperforming other models in handling complex dynamics and reflecting the data's actual characteristics. It also predicts nuanced data changes, as is evident in comparative analyzes.

\section{Conclusion}
We present KFTD, the first continuous-time two-stage framework for operational ocean spatiotemporal forecasting that tightly integrates Koopman operator theory with depth-adaptive Fourier neural operators.  By explicitly decoupling differentiable temporal interpolation from spatial prediction, KFTD eliminates the costly multi-step noise sampling endemic to diffusion-based baselines while enforcing arbitrary PDE constraints through a D-PP loss.  Extensive experiments across four variables and four hydro-dynamically distinct regions—Tropical Pacific, North Atlantic, Equatorial Indian Ocean and Southern Ocean—show consistent improvements: average MSE drops by 5.6 \% and up to 12.7 \% for SST, and seasonal 6-month forecasts retain the lowest error growth among all competing methods.  Beyond immediate forecasting skill, KFTD offers a path toward scalable, physics-aware digital twins of the global ocean: its linear Koopman latent space admits on-the-fly resolution refinement and straightforward assimilation of heterogeneous observations, while the plug-and-play residual interface allows domain scientists to incorporate new conservation laws without re-architecting the network. Future work will extend KFTD to fully coupled ocean–atmosphere prediction and explore uncertainty-aware ensemble generation for extreme-event early warning.

\section{Acknowledgments}
This work was supported in part by the National Key R\&D Program of China under Grant 2022YFB3206900, Key R\&D Program of Shandong Province of China under Grant 2023CXGC010112, the joint funds of the National Natural Science Foundation of China under Grant U24A20221, Distinguished Young Scholar of Shandong Province under Grant ZR2023JQ025, Taishan Scholars Program under Grant tstp20250708, Major Basic Research Projects of Shandong Province under Grant ZR2022ZD32.

\bibliographystyle{ACM-Reference-Format}
\balance
\bibliography{kftd}

\newpage

\end{document}